\documentclass[journal]{IEEEtran}
\usepackage{graphicx,amssymb,amsmath}
\usepackage{multicol}
\usepackage[noadjust]{cite}
\usepackage{setspace}
\usepackage{stfloats}
\usepackage{midfloat}
\usepackage[normal]{threeparttable}
\usepackage{amsthm}
\usepackage{cuted}
\usepackage{cases,subeqnarray}
\usepackage{bm,multirow,bigstrut}
\usepackage{textcomp}
\usepackage{latexsym,bm}
\usepackage{booktabs,changebar}
\usepackage{xcolor}
\usepackage{mathtools}
\usepackage{dsfont}
\usepackage{extarrows}
\newtheorem{lemm}{Lemma}

\IEEEoverridecommandlockouts
\begin{document}
\title{Achievable Rate of Rician Large-Scale MIMO Channels with Transceiver Hardware Impairments}

\author{Jiayi~Zhang,
        Linglong~Dai,
        Xinlin~Zhang,
        Emil~Bj\"{o}rnson,
        and~Zhaocheng~Wang

\thanks{Copyright $\copyright$ 2015 IEEE. Personal use of this material is permitted. However, permission to use this material for any other purposes must be obtained from the IEEE by sending a request to pubs-permissions@ieee.org.

This work was supported in part by the International Science \& Technology Cooperation Program of China (Grant No. 2015DFG12760), the National Natural Science Foundation of China (Grant Nos. 61571270 and 61201185), and China Postdoctoral Science Foundation (No. 2014M560081). The work of X. Zhang was supported in part by the Swedish Governmental Agency for Innovation Systems (VINNOVA) within the VINN Excellence Center Chase, and the Swedish Foundation for Strategic Research. E. Bj\"{o}rnson was supported by ELLIIT and the CENIIT project 15.01.}%
\thanks{J. Zhang, L. Dai and Z. Wang are with Department of Electronic Engineering as well as Tsinghua
National Laboratory of Information Science and Technology (TNList),
Tsinghua University, Beijing 100084, P. R. China (e-mails: \{jiayizhang, daill, zcwang\}@tsinghua.edu.cn).}
\thanks{X. Zhang is with Department of Signals and Systems, Chalmers University of Technology, Gothenburg, Sweden (e-mail: xinlin@chalmers.se).}
\thanks{E. Bj\"{o}rnson is with Department of Electrical Engineering (ISY), Link\"{o}ping University, Link\"{o}ping, Sweden (e-mail: emil.bjornson@liu.se).}}

\maketitle
\begin{abstract}
Transceiver hardware impairments (e.g., phase noise, in-phase/quadrature-phase (I/Q) imbalance, amplifier non-linearities, and quantization errors) have obvious degradation effects on the performance of wireless communications. While prior works have improved our knowledge on the influence of hardware impairments of single-user multiple-input multiple-output (MIMO) systems over Rayleigh fading channels, an analysis encompassing the Rician fading channel is not yet available. In this paper, we pursue a detailed analysis of regular and large-scale (LS) MIMO systems over Rician fading channels by deriving new, closed-form expressions for the achievable rate to provide several important insights for practical system design. More specifically, for regular MIMO systems with hardware impairments, there is always a finite achievable rate ceiling, which is irrespective of the transmit power and fading conditions. For LS-MIMO systems, it is interesting to find that the achievable rate loss depends on the Rician $K$-factor, which reveals that the favorable propagation in LS-MIMO systems can remove the influence of hardware impairments. However, we show that the non-ideal LS-MIMO system can still achieve high spectral efficiency due to its huge degrees of freedom.
\end{abstract}


\IEEEpeerreviewmaketitle
\section{Introduction}
By employing multiple antennas at the transceiver, wireless systems can significantly increase the spectral efficiency and transmission reliability. The capacity of single-user MIMO systems has been well investigated in the literature \cite{telatar1999capacity,zhang2015ergodic}. However, most prior works assume that ideal hardware is available at both the transmitter and receiver, which is unrealistic in practice, while the performance of practical MIMO systems is usually affected by transceiver hardware impairments, such as phase noise, I/Q imbalance, amplifier non-linearities, and quantization errors \cite{bjornson2014massive}. Although the influence of these impairments can be mitigated by calibration methods and compensation schemes at both sides, there still remains residual hardware impairments due to estimation errors, inaccurate calibration methods and different types of noise.

Recently, the large-scale (LS)-MIMO communication has drawn a substantial interest from both academia and industry as a promising technology for 5G wireless systems, such as millimeter wave (mmWave) communications. LS-MIMO systems are likely to operate in the mmWave band to accommodate many antennas within a small physical area. In LS-MIMO systems, each base station is equipped with a large number of antennas to improve the spectral and energy efficiency. Understanding the fundamental theoretical limits of the LS-MIMO system has been an active research area. For practical implementation, it is very attractive to deploy LS antenna elements with cheap, compact and power-efficient radio and digital-processing hardware. Thus, it is of profound importance to theoretically investigate how much hardware impairments can the LS-MIMO system tolerate to achieve a certain achievable rate performance.

Motivated by these observations, some researchers have analyzed the impact of transceiver hardware impairments on MIMO system performance. Specifically, experimental results to model the statistical behavior of residual hardware impairments on regular\footnote{In contrast to the LS-MIMO system, we use the terminology regular MIMO for systems with small number of antennas at the transmitter and receiver, e.g., smaller than 8 antennas.} MIMO systems have been provided in pioneering works such as \cite{studer2010mimo,studer2011system}. Utilizing this impairment model, the authors of \cite{bjornson2013capacity} and \cite{zhang2014mimo} analyzed the achievable rate of regular MIMO systems in detail. With the rapid development of LS-MIMO systems, people shift their interests to hardware impairments of LS-MIMO systems. In this context, the single type of impairments have been considered in \cite{mohammed2013per,pitarokoilis2014uplink,krishnan2015linear,Vehkapera2015asymptotic} in terms of power amplifier nonlinearities, mismatched joint decoding, and phase noise. Moreover, \cite{zhang2014mimo,Bjornson2014TIT,bjornson2014massive} examined in detail the achievable rate of LS-MIMO systems by taking into account the effects of transceiver hardware impairments.

The common characteristic of aforementioned works, however, is that they consider Rayleigh fading channels. Although the assumption of Rayleigh fading extensively simplifies the performance analysis, its validity is often violated in practical wireless propagation scenarios with the line-of-sight (LoS) path, where the Rician fading model is more general and accurate \cite{brady2013beamspace}. To the best of our knowledge, a detailed analysis of MIMO systems over Rician fading channels in the presence of transceiver hardware impairments is missing in the literature. Only recently, the high-SNR capacity limit of regular MIMO systems over Rician fading channels has been established in \cite{bjornson2013capacity}. In this paper, we aim to fill in this gap by investigating the impact of hardware impairments on the achievable rate of regular and LS-MIMO systems over Rician fading channels. Specifically, the contributions of this paper are summarized as:
\begin{itemize}
\item We derive a new analytical achievable rate expression for regular MIMO systems subject to Rician fading and hardware impairments. Although the expression is given in infinite series, the truncation error has been obtained to demonstrate its fast convergence. Additionally, we present asymptotic achievable rate expressions in the high-SNR regime, which coincide with the results of \cite{bjornson2013capacity}. Moreover, based on our analysis, there is always a ceiling on the achievable rates of regular MIMO systems.

\item For LS-MIMO systems, asymptotic expressions for the achievable rate are presented for three typical types of antenna arrays. Assuming perfect channel state information (CSI) at the receiver and no CSI at the transmitter, it is interesting to find that the achievable rate ceiling disappears by deploying a huge number of antennas at the transceiver. {Moreover, our results show that the achievable rate gap between hardware impairments and perfect hardware increases with the value of the Rician $K$-factor.}
\end{itemize}

{The remainder of the paper is organized as follows: In Section \ref{se:model}, the single-user MIMO channel model used throughout the paper is briefly introduced. Section \ref{se:ca} provides a detailed achievable rate analysis of MIMO systems with transceiver hardware impairments over Rician fading channels. A set of numerical results is given in Section \ref{se:numerical_results}. Finally, Section \ref{se:conclusion} concludes the paper.}

\section{System and Channel Model}\label{se:model}
We consider a single-user MIMO system with $N_t$ transmit antennas and $N_r$ receive antennas, and assume that perfect CSI is available at the receiver, while no CSI can be obtained at the transmitter. The system model can be written as
\begin{align}\label{eq:system}
{\bf{y }} =  {\bf{Hx }} +{\bf{n }},
\end{align}
where ${\bf{y }}\in\mathbb{C}^{N_r \times 1}$ denotes the received signal vector, ${\bf{x }}\in\mathbb{C}^{N_t \times 1}$ is the transmitted signal vector with zero mean and covariance matrix ${\tt{E}}\left[{\bf{x }}{\bf{x }}^H\right]= {\bf{Q }}$ with ${\tt{E}}[\cdot]$ being the expectation operator and $(\cdot)^H$ being the Hermitian operation, and ${\bf{n }}\in\mathbb{C}^{N_r \times 1}$ denotes the vector of zero-mean complex circularly symmetric additive white Gaussian noise (AWGN). Moreover, ${\bf{H }}\in\mathbb{C}^{N_r \times N_t}$ represents the Rician channel matrix modeling fast fading with a deterministic LoS path, which can be modelled as \cite{zhang2014power}
\begin{align}\label{eq:channel_model}
{\bf{H}} = {\sqrt{\frac{K}{K+1}}}{\bf{\bar H}} + {\sqrt{\frac{1}{K+1}}}{{\bf{H}}_\omega },
\end{align}
where {$\bar{\bf{H}}$ denotes the deterministic component,} ${{\bf{H}}_\omega }$ denotes the random fast fading component, which is composed of independent and identically distributed (i.i.d.) circularly symmetric complex Gaussian random variables with zero-mean and unit variance, and $K$ is the Rician factor denoting the power ratio between ${\bf{\bar H}}$ and ${{\bf{H}}_\omega }$. {In this paper, we normalize the channel matrix ${\bf{H}}$ as ${\tt{E}}[{\tt{tr}}({\bf{H}}{\bf{H}}^H)] = {N_r}{N_t}$, where $\tt{tr}(\cdot)$ denotes the trace of a matrix.}

In practical MIMO systems, the received signals will be unavoidably distorted by impairments of transceiver hardware components, such as filters, oscillators, converters, mixers and amplifiers, in two different ways. First, the actually emitted signals are different from the desired signals at the transmitter due to transmitter hardware impairments \cite{dai2013transceiver}. Second, the received signals may suffer from distortion after the signal processing due to receiver hardware impairments. {Although several signal compensation algorithms have been proposed and utilized at each antenna, there still remains some residual transceiver hardware impairments due to inaccurate modeling, imperfect CSI, errors in the estimation of impairments' parameters, and so forth \cite{zhang2014mimo}\footnote{{Among these residual transceiver hardware impairments, the phase noise is probably the most severe factor in single-carrier transmission, while it is still not clear in multi-carrier systems \cite{bjornson2014massive,pitarokoilis2014uplink,krishnan2015linear}.}}.} Therefore, it is important to analyze the impact of transceiver hardware impairments on the performance of MIMO systems to provide useful guidance for practical systems design.

The aggregate transceiver hardware impairments can be approximated by independent additive distortion noises at both transmitter and receiver, which has been used and verified by experiments in many previous works \cite{bjornson2014massive ,zhang2014mimo,Bjornson2014TIT}. Based on the system model \eqref{eq:system}, the actually received signal can be denoted as \cite{Bjornson2014TIT}
\begin{align}\label{eq:system_model}
{\bf{y }} =  {\bf{H}}({\bf{x}}+{\bm{\eta}_t})  + \bm{\eta}_r + {\bf{n }},
\end{align}
where the additive distortion noise terms ${\bm{\eta}_t}$ and ${\bm{\eta}_r}$ are ergodic stochastic processes that describe the hardware impairments at the transmitter and the receiver, respectively. This model is both analytically tractable, and matches experimental results accurately. The experimental results have uncovered key characteristics that ${\bm{\eta}_t}$ and ${\bm{\eta}_r}$ follow Gaussian distribution with variance proportional to the average signal power {\cite{studer2010mimo,studer2011system}}. Moreover, ${\bm{\eta}_t}$ and ${\bm{\eta}_r}$ can be analytically approximated by the central limit theorem as ${\bm{\eta}_t} \sim \mathcal{CN}(0,{\delta_t^2} \textrm{diag}(q_1,\cdots,q_{N_t}))$ and ${\bm{\eta}_r} \sim \mathcal{CN}(0,{\delta_r^2} \textrm{tr}({\bf{Q}}) {\bf{I}}_{N_r})$ \cite{Bjornson2014TIT},
where $q_1, q_2, \cdots, q_{N_t}$ are the diagonal elements of the signal covariance matrix ${\bf{Q}}$. Note that the new system model \eqref{eq:system_model} is more general than the canonical model \eqref{eq:system}, and captures dominant practical characteristics of transceiver hardware impairments. The proportionality parameters $\delta_t$ and $\delta_r$ are related to the error vector magnitude (EVM) metric, which is widely used to quantify the mismatch between the expected signal and the actual signal in RF transceivers \cite{holma2011lte}. In practical wireless systems, such as long term evolution (LTE), the EVM requirements are in the range ${\delta_t} \in [0.08,0.175]$ \cite{holma2011lte}. Note that larger values of $\delta_t$ and $\delta_r$ indicate that the MIMO system experiences higher levels of impairments caused by inaccurate transceiver hardware components. Moreover, the case of $\delta_t=\delta_r=0$ corresponds to ideal transceiver hardware components.

\section{Achievable Rate}\label{se:ca}
In this section, we present a detailed achievable rate analysis of MIMO systems with transceiver hardware impairments over Rician fading channels. Recall that neither instantaneous nor statistical CSI is available at the transmitter but perfectly known at the receiver, {we use equal power allocation on each transmit antenna as ${\bf{Q }}=\frac{P}{N_t} {\bf{I}}_{N_t}$ with the total transmit power $P$. Moreover, the average SNR per receive antenna is defined as $\rho={\tt{E}}[{\textrm{tr}}({\bf{Q}})]/N_0$, where $N_0$ denotes the noise variance, which is normalized as $N_0=1$ in the following analysis.} The new system model \eqref{eq:system_model} considering hardware impairments can be written in the form of the canonical model \eqref{eq:system} with noise variance
\begin{align}\label{eq:phi}
{\bf{\Phi }}\triangleq
\begin{cases}
\frac{{\rho \delta _t^2}}{{{N_t}}}{{\bf{H}}^H}{\bf{H}} + \left( {\rho \delta _r^2 + 1} \right){{\bf{I}}_{{N_t}}},& {\text{if}} \;\;\; {N_t} < {N_r},\\
\frac{{\rho \delta _t^2}}{{{N_t}}}{\bf{H}}{{\bf{H}}^H} + \left( {\rho \delta _r^2 + 1} \right){{\bf{I}}_{{N_r}}},& {\text{if}} \;\;\; {N_t} \geq {N_r}.
\end{cases}
\end{align}
We further assume an ergodic channel where each codeword spans over an infinite number of realizations of the fading $\bf{H}$. Then, the ergodic achievable rate $R$ can be expressed as \cite{telatar1999capacity}
\begin{align}\label{eq:ca}
{{R}}\triangleq
\begin{cases}
{{\tt{E}}}\left[ {{{\log }_2}\det \left( {{{\bf{I}}_{{N_t}}} + \frac{\rho }{{{N_t}}}{{\bf{H}}^H}{\bf{H}}{{\bf{\Phi }}^{ - 1}}} \right)} \right],& {\text{if}} \;\;\; {N_t} < {N_r},\\
{{\tt{E}}}\left[ {{{\log }_2}\det \left( {{{\bf{I}}_{{N_r}}} + \frac{\rho }{{{N_t}}}{\bf{H}}{{\bf{H}}^H}{{\bf{\Phi }}^{ - 1}}} \right)} \right],& {\text{if}} \;\;\; {N_t} \geq {N_r}.
\end{cases}
\end{align}

\newcounter{mytempeqncnt3}
\begin{figure*}[!b]
\normalsize
\setcounter{mytempeqncnt3}{\value{equation}}
\hrulefill
\vspace*{2pt}
\setcounter{equation}{13}
\begin{align}\label{eq:truncation_ca_alfano}
{R_0} &= \sum\limits_{k = {T_0}}^\infty  {\frac{{\Gamma \left( {p - q + m + k} \right)\phi _n^k}}{{\Gamma \left( {k + 1} \right)\left( {p - q + 1} \right)_k}}} \sum\limits_{t = 1}^{p - q + m + k} {\left( {{e^{\left( {K + 1} \right)/a}}{E_{p - q + m + k - t + 1}}\left( {\frac{{K + 1}}{a}} \right) - {e^{\left( {K + 1} \right)/b}}{E_{p - q + m + k - t + 1}}\left( {\frac{{K + 1}}{b}} \right)} \right)} \notag\\
 &< \sum\limits_{k = {T_0}}^\infty  {\frac{{\Gamma \left( {p - q + m + k + 1} \right)\phi _n^k}}{{\Gamma \left( {k + 1} \right)\left( {p - q + 1} \right)_k}}\left( {{e^{\left( {K + 1} \right)/a}}{E_1}\left( {\frac{{K + 1}}{a}} \right) - {e^{\left( {K + 1} \right)/b}}{E_1}\left( {\frac{{K + 1}}{b}} \right)} \right)}\notag \\
 & \xlongequal{s=k-{T_0}} \frac{{\Gamma \left( {p \!-\! q \!+\! m \!+\! {T_0} \!+\! 1} \right)\phi _n^{{T_0}}}}{{\Gamma \left( {{T_0} \!+\! 1} \right)\Gamma \left( {p \!-\! q \!+\! {T_0} \!+\! 1} \right)}}\sum\limits_{s = 0}^\infty  \frac{{{{\left( {p \!-\! q \!+\! m \!+\! {T_0} \!+\! 1} \right)}_s}{{\left( 1 \right)}_s}}}{{{{\left( {{T_0} \!+\! 1} \right)}_s}{{\left( {p \!-\! q \!+\! {T_0} \!+\! 1} \right)}_s}}}\frac{{\phi _n^s}}{{s!}} \left( {{e^{\left( {K \!+\! 1} \right)/a}}{E_1}\left( {\frac{{K \!+\! 1}}{a}} \right) \!-\! {e^{\left( {K \!+\! 1} \right)/b}}{E_1}\left( {\frac{{K \!+\! 1}}{b}} \right)} \right) \notag \\
& = \frac{{\Gamma \left( {p - q + m + {T_0} + 1} \right)\phi _n^{{T_0}}}}{{\Gamma \left( {{T_0} + 1} \right) \left( {p - q + 1} \right)_{T_0}}}{}_2{F_2}\left( {p - q + m + {T_0} + 1,1;{T_0} + 1,p - q + {T_0} + 1;\phi _n} \right)\notag \\
&\quad\times \left( {{e^{\left( {K + 1} \right)/a}}{E_1}\left( {\frac{{K + 1}}{a}} \right) - {e^{\left( {K + 1} \right)/b}}{E_1}\left( {\frac{{K + 1}}{b}} \right)} \right),
\end{align}
\setcounter{equation}{\value{mytempeqncnt3}}
\end{figure*}

\subsection{Exact Analysis}
For notational convenience, we define $p \triangleq \max(N_t,N_r)$, $q \triangleq \min(N_t,N_r)$, and the instantaneous MIMO channel correlation matrix ${\bf{W}}$ as
\begin{align}\label{eq:W}
{\bf{W}}\triangleq
\begin{cases}
{\bf{H}}^H{\bf{H}},& {\text{if}} \;\;\; {N_t} < {N_r},\\
{\bf{H}}{{\bf{H}}^H},& {\text{if}} \;\;\; {N_t} \geq {N_r}.
\end{cases}
\end{align}
Note that ${\bf{W}}$ is a complex non-central Wishart matrix \cite{zanella2009marginal}.
{\begin{lemm}\label{lemm:exact_rate}
The exact achievable rate of MIMO systems with residual hardware impairments over Rician fading channels can be expressed as
\begin{align}\label{eq:exact_ca_alfano}
{R} &= \frac{{G}}{{\ln (2)}}\sum\limits_{n = 1}^q {\sum\limits_{m = 1}^q {{D_{n,m}}} } \sum\limits_{k = 0}^\infty  {\frac{{\Gamma \left( {p - q + m + k} \right)\phi  _n^k}}{{\Gamma \left( {k + 1} \right)  \left( {p - q + 1} \right)_k}}}  \notag \\
&\times \sum\limits_{t = 1}^{p - q + m + k} \Bigg( {e^{\left( {K + 1} \right)/a}}{E_{p - q + m + k - t + 1}}\left( {\frac{{K + 1}}{a}} \right) \notag \\
& \quad - {e^{\left( {K + 1} \right)/b}}{E_{p - q + m + k - t + 1}}\left( {\frac{{K + 1}}{b}} \right) \Bigg),
\end{align}
where $(x)_z\triangleq\Gamma(x+z)/\Gamma(x)$, $a  \triangleq \frac{{\rho \left( {1 + \delta _t^2} \right)}}{{{N_t}\left( {1 + \rho \delta _r^2} \right)}}$, $b  \triangleq \frac{{\rho \delta _t^2}}{{{N_t}\left( {1 + \rho \delta _r^2} \right)}}$, ${E_z}\left( x \right) = \int_1^\infty  {{t^{ - z}}} {e^{ - xt}}dt$ is the exponential integral function \cite[Eq. (8.211.1)]{gradshtein2000table}, ${\bm{\phi }}=[\phi_1, \phi_2,\cdots, \phi_q]^T$ is the squared singular values of ${\sqrt{K}\bar{\bf{H}}}$, and
\begin{align}
{G} &\triangleq \frac{{\prod\nolimits_{i = 1}^q {{e^{ - {\phi _i}}}} }}{{{{\left[ {\left( {p - q} \right)!} \right]}^q}\prod\limits_{1 \le i < j \le q} {\left( {{\phi _j} - {\phi _i}} \right)}} }.
\end{align}
Moreover, $D_{n,m}$ denotes the $(n,m)$th cofactor of the $(q\times q)$ matrix $\bf{\Omega }$, whose elements are given by
\begin{align}\label{eq:Omega_alfano}
{{\bf{\Omega }} _{n,m}} = \Gamma \left( {p - q +  m} \right){}_1{F_1}\left( {p - q + m,p - q + 1,\phi_n } \right),
\end{align}
where ${}_1{F_1}\left( \cdot \right)$ is the confluent hypergeometric function \cite[Eq. (9.21)]{gradshtein2000table}.
\end{lemm}}
{\begin{IEEEproof}
The marginal probability density function (PDF) of an unordered squared singular value of $\bf{W}$ is given by \cite{alfano2004mutual}
\begin{align}\label{eq:PDF_alfano}
f\left( \lambda  \right) &= \frac{{G{e^{ - \lambda \left( {K + 1} \right)}}}}{{q\lambda }}\sum\limits_{n = 1}^q \sum\limits_{m = 1}^q {{D_{n,m}}}{{\left( {\left( {K + 1} \right)\lambda } \right)}^{p - q + m}} \notag \\
 &\times {}_0{F_1}\left( {p - q + 1;\left( {K + 1} \right){\phi _n}\lambda } \right),
\end{align}
where ${}_0{F_1}(\cdot)$ denotes the hypergeometric functions \cite[Eq. (9.14)]{gradshtein2000table} and can be expressed as ${}_0{F_1}\left( {x,y} \right) = \sum\limits_{m = 0}^\infty  {\frac{{{y^m}}}{{m!{{\left( x \right)}_m}}}}$ \cite{abramowitz1964handbook}.
We can rewrite \eqref{eq:ca} as
\begin{align}\label{eq:ca_extend}
R &= \frac{q}{{\ln 2}}{\tt{E}}\left[ {\ln \left( {1 + \frac{{\rho \lambda /{N_t}}}{{\rho \delta _t^2\lambda /{N_t} + \rho \delta _r^2 + 1}}} \right)} \right]\notag \\
&= \frac{q}{{\ln 2}}\left({\tt{E}}\left[ \ln \left( {1 + a\lambda } \right)\right] - {\tt{E}}\left[\ln \left( {1 + b\lambda } \right) \right]\right).
\end{align}
By substituting \eqref{eq:PDF_alfano} into \eqref{eq:ca_extend}, the first expectation of \eqref{eq:ca_extend} can be derived as
\begin{align}\label{eq:ca_extend_1}
&{\tt{E}}\left[ {\ln \left( {1 + a\lambda } \right)} \right] = \int_0^\infty  \ln \left( {1 + a\lambda } \right)\frac{{G{e^{ - \lambda \left( {K + 1} \right)}}}}{{q\lambda }}\sum\limits_{n = 1}^q \sum\limits_{m = 1}^q {{D_{n,m}}} \notag \\
& \times {}_0{F_1}\left( {p - q + 1;\left( {K + 1} \right){\phi _n}\lambda } \right){{\left( {\left( {K + 1} \right)\lambda } \right)}^{p - q + m}} d\lambda \notag \\
&= \frac{G}{q}\sum\limits_{n = 1}^q {\sum\limits_{m = 1}^q {{D_{n,m}}} \sum\limits_{k = 0}^\infty  {\frac{{\Gamma \left( {p - q + m + k} \right)\phi _n^k}}{{\Gamma \left( {k + 1} \right){{\left( {p - q + 1} \right)}_k}}}} }\notag \\
&  \times\sum\limits_{t = 1}^{p - q + m + k} {{e^{\left( {K + 1} \right)/a}}{E_{p - q + m + k - t + 1}}\left( {\frac{{K + 1}}{\alpha }} \right)},
\end{align}
where we have used the following integral identity \cite{alfano2004mutual}
\begin{align}\label{eq:integral}
\int_0^\infty  {\ln \left( {1 \!+\! \alpha x} \right)}  \frac{{{{x^{z  \!-\! 1}}}}}{{{e^{\beta x}}}}dx = \frac{{{\Gamma \left( z  \right)}}{{{e^{  \beta/\alpha }}}}} {\beta^{z}}\sum\limits_{l = 1}^z  {{E_{z  \!-\! l \!+\! 1}}\left( {\frac{\beta}{\alpha }} \right)}.
\end{align}
The second expectation of \eqref{eq:ca_extend} can be derived in a similar way. Then, the proof is ended by substituting the corresponding results (e.g.,  \eqref{eq:ca_extend_1}) into \eqref{eq:ca_extend}.
\end{IEEEproof}}

To show the fast convergence of the infinite series in \eqref{eq:exact_ca_alfano}, we assume that only the $T_0-1$ first terms are used. Note that if $x<y$, the function ${e^x}E_n(x)-{e^y}E_n(y)$ is monotonically decreasing in $n$ according to the derivative property of $E_n(x)$ \cite[Eq. (5.1.26)]{abramowitz1964handbook}. {Then the truncation error $R_0$ is upper bounded as \eqref{eq:truncation_ca_alfano} at the bottom of next page,
where ${}_2{F_2}\left( {{\alpha _1},{\alpha _2};{\beta _1},{\beta _2};z} \right) = \sum\limits_{k = 0}^\infty  {\frac{{{{\left( {{\alpha _1}} \right)}_k}{{\left( {{\alpha _2}} \right)}_k}}}{{{{\left( {{\beta _1}} \right)}_k}{{\left( {{\beta _2}} \right)}_k}}}\frac{{{z^k}}}{{k!}}} $ is the generalized hypergeometric function \cite[Eq. (9.14.1)]{gradshtein2000table}.} Moreover, the required terms of series $T_0$ has been investigated in Table \ref{table1} for different parameters. To achieve a satisfactory accuracy, e.g., $10^{-6}$, more terms are needed for larger values of $K$, $N_t$ and $N_r$. On the contrary, $T_0$ decreases with the larger values of SNR $\rho$. Finally, for all cases considered in Table \ref{table1}, only less than $15$ terms need to be calculated.

\begin{table}[!t]
\renewcommand{\thetable}{\Roman{table}}
\caption{Required terms of series $T_0$ to achieve a satisfactory accuracy ($\leq 10^{-6}$ )}
\label{table1}
\centering
\begin{tabular}{ccccccc}
\toprule[2pt]
  $\rho$  & $N_t$ & $N_r$   & $\delta_t$ & $\delta_r$  & $K$ & $T_0$\\
\midrule[1pt]
  0     &  2      &   2 &   0.15&   0.15&   1&   11 \\
  0     &  2      &   2 &   0.15&   0.15&   5&   15 \\
  10     &  2      &   2 &   0.15&   0.15&   1&   10 \\
  0     &  4      &   4 &   0.15&   0.15&   1&   12  \\
  0     &  2      &   2 &   0.1&   0.1&   1&    12 \\
 \bottomrule[2pt]
\end{tabular}
\end{table}

\subsection{High-SNR Analysis}\label{se:high}
Although \eqref{eq:exact_ca_alfano} is the exact achievable rate, it provides little insight on how hardware impairments affect the achievable rate of MIMO systems over Rician fading channels. For high-SNR values, we can take $\rho \rightarrow \infty$ in \eqref{eq:ca}, and follow a similar line of reasoning as
in Lemma \ref{lemm:exact_rate}. Then, the asymptotic achievable rate approaches the finite limit
\setcounter{equation}{14}
\begin{align}\label{eq:high_ca}
{R}^\infty &=\frac{{G}}{{\ln 2}}\sum\limits_{n = 1}^q {\sum\limits_{m = 1}^q {{D_{n,m}}} } \sum\limits_{k = 0}^\infty  {\frac{{\Gamma \left( {p - q + m + k} \right)\phi  _n^k}}{{\Gamma \left( {k + 1} \right)  \left( {p - q   + 1} \right)_k}}}\notag \\
& \times \sum\limits_{t = 1}^{p - q + m + k} \Bigg({e^{\left( {K + 1} \right)/{a^{'}}}}{E_{p - q + m + k - t + 1}}\left( {\frac{{K + 1}}{{a^{'}}}} \right)  \notag \\
&\quad - {e^{\left( {K + 1} \right)/{b^{'}}}}{E_{p - q + m + k - t + 1}}\left( {\frac{{K + 1}}{{b^{'}}}} \right) \Bigg),
\end{align}
where ${a^{'}}  \triangleq \frac{{  \left( {1 + \delta _t^2} \right)}}{{{N_t}\delta _r^2}}$ and $b^{'} \triangleq \frac{{ \delta _t^2}}{{{N_t}  { \delta _r^2}  }}$, respectively.

The term $\left( {{e^{\left( {K \!+\! 1} \right)/{a^{'}}}}{E_1}\left( {\frac{{K \!+\! 1}}{{a^{'}}}} \right) \!-\! {e^{\left( {K \!+\! 1} \right)/{b^{'}}}}{E_1}\left( {\frac{{K \!+\! 1}}{b^{'}}} \right)} \right)$ in \eqref{eq:high_ca} becomes zero when $k$ is large \cite{gradshtein2000table}. Therefore, the achievable rate of MIMO systems over Rician fading channels with residual hardware impairments approaches a finite ceiling in the high-SNR regime, which is also found in the case of Rayleigh fading channels in \cite{zhang2014mimo} and the case of any fading channels with only transmitter impairments in \cite{bjornson2013capacity}. This effect can be explained as that the transceiver distortion will increase with the transmit power. Accordingly, the equivalent SNR, $\frac{\rho }{{{N_t}}}{\bf{H}}{{\bf{H}}^H}{{\bf{\Phi }}^{ - 1}}$, in \eqref{eq:ca} will not increase. However, the achievable rate $R$ will increase to infinity with SNR if adopting the ideal hardware. Moreover, \eqref{eq:high_ca} reveals that the residual hardware impairments dominate on the achievable rate performance of MIMO systems in the high-SNR regime.

Moreover, assuming that the first $T_0^{'}-1$ terms are used in the infinite series, the truncation error $R_0^{'}$ is upper bounded as
\begin{align}\label{eq:truncation_ca_alfano_high}
& {R_0^{'}} \le \frac{{\Gamma \left( {p - q + m + T_0^{'} + 1} \right)\phi _n^{{T_0^{'}}}}}{{\Gamma \left( {T_0^{'} + 1} \right)\left( {p - q   + 1} \right)_{T_0^{'}}}}\notag \\
& \times{}_2{F_2}\left( {p \!-\! q \!+\! m \!+\! {T_0^{'}} \!+\! 1,1;{T_0^{'}} \!+\! 1,p \!-\! q \!+\! {T_0^{'}} \!+\! 1;\phi _n} \right)\notag \\
& \times \left( {{e^{\left( {K \!+\! 1} \right)/{a^{'}}}}{E_1}\left( {\frac{{K \!+\! 1}}{{a^{'}}}} \right) \!-\! {e^{\left( {K \!+\! 1} \right)/{b^{'}}}}{E_1}\left( {\frac{{K \!+\! 1}}{b^{'}}} \right)} \right).
\end{align}

\subsection{Asymptotic LS-MIMO Analysis}
In the following, we consider the achievable rate of three asymptotic antenna deployment in LS-MIMO systems. Note that our analysis holds for any LoS model that satisfy the limit of ${\frac{1 }{{{p}}}{\bf{H}}{{\bf{H}}^H}} \xrightarrow {a.s.}{{\bf{I}}_{{q}}}$. If a uniform linear array (ULA) is adopted at the transmitter, the $({m,n})$th entry ${\bf{\bar H}}_{mn} $ is given by
\begin{align}\label{eq:Los_model}
{\bf{\bar H}}_{mn} = e^{-j(m-1) (2\pi d/\lambda) \sin \theta_{n}},
\end{align}
where $d$ is the transmit antenna spacing, $\lambda$ is the wavelength, and $\theta_{n}$ is
the arrival angle of the $n$th receive antenna. Moreover, we set $d=\lambda/2$, which means that there is no correlation between receive antennas.

First, the number of transmit antennas $N_t$ tends to infinity while the number of receiver antennas $N_r$ is fixed. According to the law of large numbers, the correlation matrix ${\frac{1 }{{{N_t}}}{\bf{H}}{{\bf{H}}^H}} -{{\bf{I}}_{{N_r}}} \xrightarrow {a.s.} {\bf{0}}$ \cite[Lemma 2]{zhang2014power} as $N_t \rightarrow \infty$, where $a.s.$ denotes almost sure convergence. To take the limit inside the expectation in \eqref{eq:ca} by the dominated convergence theorem \cite{couillet2011random}, the achievable rate reduces to
\begin{align}\label{eq:nt_ca}
{R_{{N_t} \to \infty }} = {N_r}{\log _2}\left( {1 + \frac{\rho }{{\rho \delta _t^2 + \rho \delta _r^2 + 1}}} \right),
\end{align}
which indicates that the achievable rate of LS-MIMO systems with infinite $N_t$ depends on the transceiver distortions, transmit SNR and the number of receiver antennas $N_r$. Moreover, as we increase $N_r$, the achievable rate grows linearly. However, if $N_r$ is fixed but SNR is increased, the achievable rate asymptotically approaches the limit as we discuss in Section \ref{se:high}. This fact suggests that the achievable rate will saturate in the high-SNR regime for Rician fading channels.

Then, we consider the second case, where the receiver employs large number of receiver antennas $N_r$ but the number of transmit antennas $N_t$ is fixed. Recall that the ULA model is assumed and multiplying the term of $\left( {{{\bf{I}}_{{N_t}}} + \frac{\rho }{{{N_t}}}{{\bf{H}}^H}{\bf{H}}{{\bf{\Phi }}^{ - 1}}} \right)$ in \eqref{eq:ca} by $1/N_r$, the achievable rate \eqref{eq:ca} can be written as
\begin{align}\label{eq:nr_ca}
R = {\tt{E}}\left\{ {{{\log }_2}\det \left( {{{\bf{I}}_{{N_t}}} + \frac{{\frac{\rho }{{{N_t}{N_r}}}{{\bf{H}}^H}{\bf{H}}}}{{\frac{{\rho \delta _t^2}}{{{N_t}{N_r}}}{{\bf{H}}^H}{\bf{H}} + \frac{{\left( {\rho \delta _r^2 + 1} \right)}}{{{N_r}}}{{\bf{I}}_{{N_t}}}}}} \right)} \right\}.
\end{align}
As $N_r \rightarrow \infty$, we utilize the dominated convergence theorem and the fact that the noise term and receiver distortion term go to zero. Then, \eqref{eq:nr_ca} can approach to
\begin{align}\label{eq:nr_ca_1}
{{R}_{{N_r} \to \infty }}   = {N_t}{\log _2}\left( {1 + \frac{1}{{\delta _t^2}}} \right).
\end{align}


It is clear that the achievable rate grows linearly with the number of transmit antennas $N_t$. Moreover, the receiver distortion (denoted by $\delta_r^2$) and the SNR have no impact on the achievable rate performance. This shows the key difference from the first case of large $N_t$ but fixed $N_r$, where both transceiver distortions (denoted by $\delta_t^2$ and $\delta_r^2$) characterize the system achievable rate performance. Such result suggests that employing low-cost hardware at the receiver is suitable if hardware impairments are unavoidable.

Finally, the third case of large $N_t$ and $N_r$, and general Rician fading model are considered, where the achievable rate ${R }$ can be reexpressed as
{
\begin{align}\label{eq:nt_nr_ca}
{R  }&= {\tt{E}}\Bigg[ {{\log }_2}\det \left( {\frac{{\rho \left( {1 + \delta _t^2} \right)}}{{{N_t}}}{\bf{H}}{{\bf{H}}^H} + \left( {\rho \delta _r^2 + 1} \right){{\bf{I}}_{{N_r}}}} \right) \notag \\
&- {{\log }_2}\det \left( {\frac{{\rho \delta _t^2}}{{{N_t}}}{\bf{H}}{{\bf{H}}^H} + \left( {\rho \delta _r^2 + 1} \right){{\bf{I}}_{{N_r}}}} \right) \Bigg]\notag \\
&  = {\tt{E}}\Bigg[ {{\log }_2}\det \left( { a{\bf{H}}{{\bf{H}}^H} + {{\bf{I}}_{{N_r}}}} \right) + {N_r}{{\log }_2}\left( {\rho \delta _r^2 + 1} \right) \notag \\
&- {{\log }_2}\det \left( {b{\bf{H}}{{\bf{H}}^H} + {{\bf{I}}_{{N_r}}}} \right) - {N_r}{{\log }_2}\left( {\rho \delta _r^2 + 1} \right) \Bigg]\notag \\
&  = {\tt{E}}\left[ {{{\log }_2}\det \left( {a{\bf{H}}{{\bf{H}}^H} + {{\bf{I}}_{{N_r}}}} \right) - {{\log }_2}\det \left( {b{\bf{H}}{{\bf{H}}^H} + {{\bf{I}}_{{N_r}}}} \right)} \right] \notag \\
&  = J(1/a, {{\bf{I}}_{{N_r}}}) - J(1/b, {{\bf{I}}_{{N_r}}}),
\end{align}}
where $J(1/a, {{\bf{I}}_{{N_r}}}) \triangleq E\left[ {{{\log }_2}\det \left( {a{\bf{H}}{{\bf{H}}^H} + {{\bf{I}}_{{N_r}}}} \right)}\right]$ and $J(1/b, {{\bf{I}}_{{N_r}}}) \triangleq E\left[ {{{\log }_2}\det \left( {b{\bf{H}}{{\bf{H}}^H} + {{\bf{I}}_{{N_r}}}} \right)}\right]$, respectively. From \cite[Theorem 6.14]{couillet2011random}, we have a large-system approximation of the achievable rate $J(1/a, {{\bf{I}}_{{N_r}}})$ for a large number of antennas at both transmitter and receiver sides ($N_t, N_r \rightarrow \infty$) and uniform transmit power allocation as \cite[Eq. (13.10)]{couillet2011random}
\begin{align}\label{eq:nt_nr_ca_ap}
&J(1/a, {{\bf{I}}_{{N_r}}}) - \Bigg[ {{\log }_2}\det \left( { {{a{{\bf{\Psi }}^{ - 1}}+ {\bf{{\bar H}\bar \Psi }}{{\bf{{\bar H}}}^T}}}} \right)  \notag \\
&  + {{\log }_2}\det \left( {a}{{\bf{\Psi }}^{ - 1}} \right) - \frac{{{{\log }_2}\left( e \right)}}{{{a N_t}}}\sum\limits_{i,j} {{\frac{{v_i}{{\bar v}_j}}{K+1}}}  \Bigg]\xrightarrow {a.s.} {{0}},
\end{align}
where $\bf{\Psi }$ denotes the diagonal matrix with the $i$th entry ${ {\psi }}_i$, and $\bf{\bar \Psi }$ is the diagonal matrix with the $j$th entry ${ {\bar \psi }}_j$, respectively. Moreover, we define $v_i$ and ${\bar v}_j$ as the $i$th diagonal entry of $\left({{{\bf{\bar \Psi }}}^{ - 1}} + \frac{1}{a}{{\bf{{\bar H}}}^T}{\bf{\Psi {\bar H}}}\right)^{-1}$ and the $j$th diagonal entry of $\left({{\bf{\Psi }}^{ - 1}} + \frac{1}{a}{\bf{{\bar H}\bar \Psi }}{{\bf{{\bar H}}}^T}\right)^{-1}$, respectively. As $N_t \rightarrow \infty$, the error between the right hand side of \eqref{eq:nt_nr_ca_ap} goes almost sure to zero.  It is clear from \eqref{eq:nt_nr_ca_ap} that the approximation error decreases asymptotically by increasing the number of transmit antennas $N_t$. The entries ${{\psi }}_i$ and ${{\bar \psi }}_j$ can be obtained by solving the following equations as
\begin{equation}\label{eq:nt_nr_ca_unique_iid}
  \left\{\begin{aligned}
\psi  &= a{\left[ {1 + \frac{1}{{{N_t}(K+1)}}{\tt{tr}}\left\{{{\left( {\frac{1}{{\bar \psi }} {{\bf{I}}_{N_t}} + \frac{\psi}{a} {{\bf{\bar H}}^T}{\bf{\bar H}}} \right)}^{ - 1}}\right\}} \right]^{ - 1}} , \\
{\bar \psi }  &= a{\left[ {1 + \frac{1}{{{N_t}(K+1)}}{\tt{tr}}\left\{{{\left( {\frac{1}{\psi }{{\bf{I}}_{N_r}} + \frac{\bar \psi}{a} {\bf{\bar H}}{{\bf{\bar H}}^T}} \right)}^{ - 1}}\right\}} \right]^{ - 1}}.
\end{aligned}
  \right.
\end{equation}

Note that equations in \eqref{eq:nt_nr_ca_unique_iid} are fixed point iterations. The unknown variables $\psi$ and $\bar \psi$ can be easily obtained by solving the formulas in \eqref{eq:nt_nr_ca_unique_iid}. Substituting $\psi$ and $\bar \psi$ into \eqref{eq:nt_nr_ca_ap} and using the similar method to calculate $J(1/b, {{\bf{I}}_{{N_r}}})$, the desired achievable rate in \eqref{eq:nt_nr_ca} can be derived.

\section{Numerical Results}\label{se:numerical_results}
In this section, we illustrate the key analytical insights presented in Section \ref{se:ca} by various Monte-Carlo simulations. For the ideal and non-ideal system, the achievable rate results have been obtained by means of Monte-Carlo simulations using $10^{6}$ trails, respectively. Furthermore, the LoS model in \eqref{eq:Los_model} has been used in our simulations.


\begin{figure}[tb]
\centering
\includegraphics[scale=0.65]{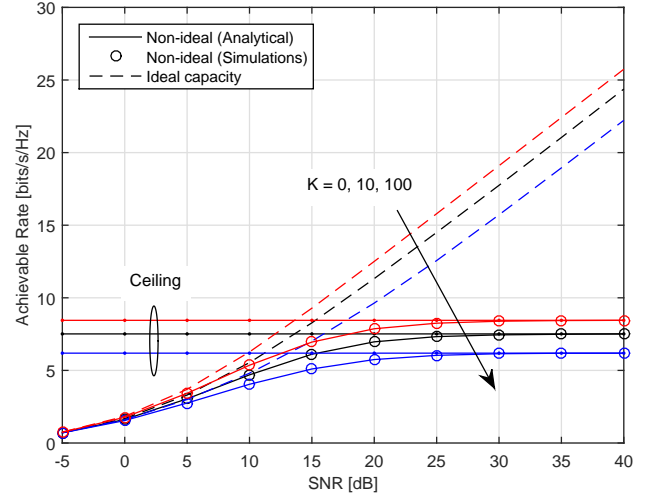}\\
\caption{Achievable rate of regular MIMO systems with hardware impairments against SNR and Rician $K$-factor, where $\delta_t =\delta_r= 0.15$ and $N_t=N_r=2$.
\label{fig:small_Nt_Nr}}
\end{figure}

\begin{figure}[tb]
\centering
\includegraphics[scale=0.65]{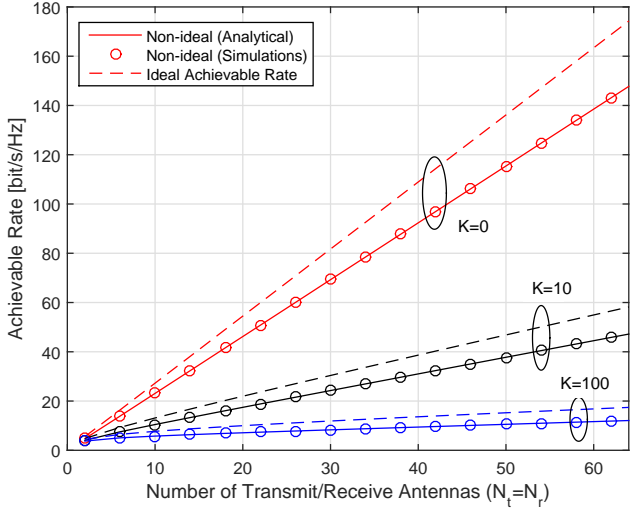}\\
\caption{Achievable rate of LS-MIMO systems with hardware impairments against the number of transmit and receive antennas and Rician $K$-factor, where $\delta_t =\delta_r= 0.15$, $\rho = 10$dB, and $N_t=N_r$.
\label{fig:large_Nt_Nr}}
\end{figure}

\begin{figure}[tb]
\centering
\includegraphics[scale=0.65]{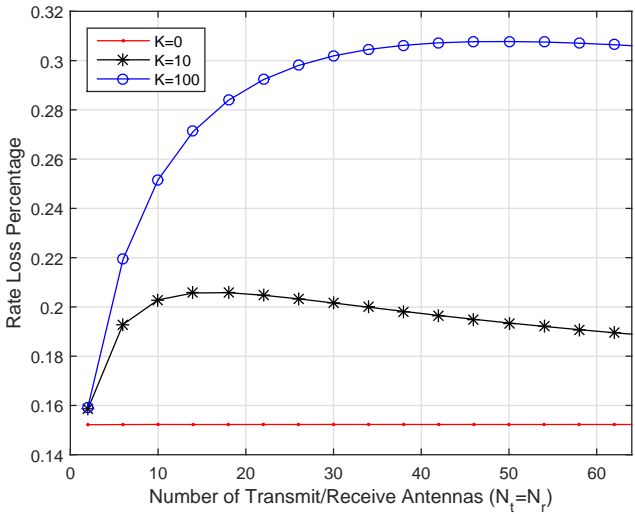}\\
\caption{Achievable rate loss of LS-MIMO systems with hardware impairments against the number of transmit antennas $N_t$ and Rician $K$-factor, where $\delta_t =\delta_r= 0.15$, $\rho = 10$dB, and $N_t=N_r$.
\label{fig:C_large_Nt_Nr_diff_K_64_loss}}
\end{figure}

In Fig. \ref{fig:small_Nt_Nr}, the simulated achievable rate, the analytical result \eqref{eq:exact_ca_alfano} and the high-SNR approximation \eqref{eq:high_ca} of regular MIMO systems with hardware impairments are plotted against the SNR and Rician $K$-factor, where $\delta_t =\delta_r= 0.15$ and $N_t=N_r=2$ are considered. Figure 1 validates the accuracy of our derived analytical expressions in \eqref{eq:exact_ca_alfano} and \eqref{eq:high_ca}. For the case of hardware impairments, it is clear that there is a finite rate ceiling, which cannot be crossed by increasing the SNR value. Furthermore, we observe that an increase in SNR tends to increase the achievable rate of both ideal and non-ideal system, albeit the relative difference between the curves gets steadily larger. In addition, a higher $K$ value yields lower achievable rate, although the gap between the corresponding curves decreases as $K$ increases, which implies that its effect becomes less pronounced.

The achievable rate of single-user LS-MIMO systems with ideal and non-ideal hardware is shown in Fig. \ref{fig:large_Nt_Nr}, which reveals that the finite achievable rate ceiling disappears for large numbers of transmit and receive antennas. This phenomenon is consistent with the results with Rayleigh fading channels in \cite{zhang2014mimo}. This is because the reduction in effective SNR, $\rho/N_t$, can be compensated by the large array gain at the receiver. As expected, the increased LoS component (larger values of $K$) will decrease the rank of the correlation matrix and the system's achievable rate.

To further investigate the effect of the Rician $K$-factor on the achievable rate of LS-MIMO systems, we introduce a new metric as $R_{\tt{loss}}={(R_{\tt{ideal}}-R_{\tt{non-ideal}})/R_{\tt{ideal}}}$, which denotes the achievable rate loss between ideal and non-ideal system with hardware impairments. Moreover, we assume that the number of transmit and receive antennas grows together. It is important to observe from Fig. \ref{fig:C_large_Nt_Nr_diff_K_64_loss} that the achievable rate loss $R_{\tt{loss}}$ increases with the value of the Rician $K$-factor. However, with a relatively large number of antennas at both transmitter and receiver sides, the achievable rate loss approaches a finite value. For example, the relative achievable rate loss $R_{\tt{loss}}$ for $K=0$ is around 15\%, while $R_{\tt{loss}} \rightarrow 30.5\%$ for the case of $K=100$. Therefore, it is more important to utilize ideal hardware at LS-MIMO systems when operating over strong LoS environment.

\section{Conclusions}\label{se:conclusion}
In this paper, we present a detail achievable rate analysis of regular and LS-MIMO systems under transceiver hardware impairments and Rician fading conditions. New analytical achievable rate results are derived for finite and infinite number of transceiver antennas. We obtain an asymptotic high-SNR achievable rate expression to reveal a finite ceiling in regular MIMO systems. Moreover, the impact of the Rician $K$-factor and hardware impairments on the achievable rate performance are investigated. Our findings reveal that the achievable rate ceiling vanishes by increasing both the number of transmit and receive antennas in LS-MIMO systems. Finally, we conclude that the achievable rate loss due to hardware impairments increases with the value of the Rician $K$-factor.


\bibliographystyle{IEEEtran}
\bibliography{IEEEabrv,Ref}

\end{document}